# THE STAR OF BETHLEHEM IS *NOT* THE NOVA DO AQUILAE (NOR ANY OTHER NOVA, SUPERNOVA, OR COMET)

*By Bradley E. Schaefer*
*Department of Physics and Astronomy, Louisiana State University*

  The Star of Bethlehem is only known from a few verses in the Gospel of Matthew, with the Star inspiring and leading the Magi (i.e., Persian astrologers) to Jerusalem and ultimately worshipping the young Jesus Christ in Bethlehem.  In the last four centuries, astronomers have put forth over a dozen greatly different naturalistic explanations, all involving astronomical events, often a bright nova, supernova, or comet.  This paper will evaluate one prominent recent proposal, that the Star was a 'recurrent nova' now catalogued as DO Aquilae, and provide three refutations.  In particular, (1) DO Aql is certainly not a recurrent nova, but rather an ordinary nova with a recurrence time scale of over a million years, (2) in its 1925 eruption, DO Aql certainly never got brighter than 8.5 mag, and the physics of the system proves that it could never get to the required luminosity of a supernova, and (3) the Magi were astrologers who had no recognition or interpretation for novae (or supernovae or comets) so any such event is completely irrelevant and meaningless to them.

*Introduction*

    In a 1999 book titled "*The Star of Bethlehem, an Astronomer's View*" by Mark Kidger[1], he started with a standard recounting of the story and reasonably concluded that the birth of Jesus (and hence the Star) was in the springtime of 5 B.C. or 6 B.C.  He then focused on  separate Chinese reports of two comet-like 'guest stars' from 5 B.C. and from 4 B.C., finally concluding that these were both reports of a single *nova* eruption that appeared between Altair and the head of Capricorn in the spring of 5 BC .  Then, based on the date of the alleged nova, he connected it to the Star of Bethlehem with the only logic being "Just by the coincidences in dates, the 5 B.C. object was -- *must* have been -- the Star."  (His italics.)  On looking through the catalogue of variable stars, the closest nova is DO Aql (Nova Aquilae 1925), so he supposed that DO Aql is a recurrent nova with an eruption roughly two millennia earlier.

*The First Refutation; DO Aql Is Not A Recurrent Nova*

Kidger's hypothesis requires that the Star erupted in 5 B.C. (to be reported by the Chinese as a concatenation of a "broom star" in 5 B.C. and a "fuzzy star" in 4 B.C.) as well as in 1925 (being catalogued as the ordinary nova DO Aql), and he calls the system as a recurrent nova. Kidger offers no evidence to show that DO Aql is a recurrent nova, and he says only "it is definitely easier to limit the search to recent novas, just in case one of them happened to be a repeat outburst of the putatitve 5 B.C. nova explosion." However, there are strong reasons for knowing that DO Aql cannot be a recurrent nova, and that it has a near-maximal recurrence time scale.

In an exhaustive compilation of recurrent nova properties[2] and nova properties[3], Ashley Pagnotta and I have recognized and quantified seven observable properties that can be used to distinguish recurrent novae (with eruptions separated by historical time scales) from classical novae with long recurrence time scales (with recurrence time scales of 10,000 to over a million years). Some nova systems might be recurrent with only one of the multiple eruptions in historic times being discovered, so our criteria for recurrence is useful to recognize such cases. For the particular case of DO Aql, we have observations that can test only four of our criteria. The first property is that only low excitation lines were seen in spectra of DO Aql[4] (e.g., the iron lines go no higher than Fe III), whereas *all* recurrent novae show very high excitation lines (i.e., He II and Fe VII to Fe XIV). The second property is the amplitude and decline rate, with DO Aql having an amplitude of 9.5 mag and the time to fade by three magnitudes from peak (called $t_3$) being 900 days. With the very long fade time, DO Aql lies on the edge of the region for novae, far away from the recurrent novae. The third property is the presence of an evolved companion star, as indicated by an orbital period longer than 0.6 days, with 80% of the recurrent novae having evolved companions. DO Aql has an orbital period of 0.168 days[5], and thus it certainly does not have an evolved companion. The fourth property is the flat top in the peak of the light curve, which is F-class for DO Aql, whereas all known recurrent novae have either S-class or P-class. Thus, all individual properties of DO Aql point consistently to properties that are the complete opposite of those for recurrent novae.

In my massive compilation and classification of nova light curves[6], I had adopted DO Aql as the prototype of the "F" class novae, those with a long flat peak at nearly constant luminosity. The F-class novae are defined by their very long time at peak, with this requiring continuing nuclear burning on the surface of the white dwarf (which points to a very low mass white dwarf that must have a very long recurrence time scale[7]). In addition, the long ejection of material throughout the peak makes for a high ejected mass such as requires a long recurrence time scale to accumulate enough material. (I have measured a large ejected mass for the F-class nova BT Mon by means of the orbital period change across its 1939 eruption[8].) Detailed models[9] of the long-duration peaks require low mass white dwarfs, ~0.6 $M_o$, for which the recurrence time scale must be at its maximum[7]. In addition, the F-class novae have low-energy emission lines, with this pointing to a low mass white dwarf (that would have a very long recurrence time scale). Thus, with good confidence, we know that DO Aql and all F-class novae have the maximal recurrence time scale, over a million years, and so DO Aql is not a recurrent nova and did not go off roughly 2000 years ago.

*The Second Refutation; DO Aql Never Came Near to Naked Eye Visibility*

I have already published the full light curve of the 1925 eruption as based on magnitudes reported widely in the literature[6], with this being entirely in the V-band. In late 2009, I used the SMARTS 1.3-m telescope on Cerro Tololo to measure late-time quiescent magnitudes of B=18.55 and 18.63 for B-V=0.56 and 0.59 respectively. In late 2010, I traveled to the Harvard College Observatory and measured the B-band light curve with their archival plates from 1899 to 1934. The first Palomar Sky Survey, in the 1950s shows DO Aql at B=18.10, while the POSS2 survey shows B=18.80. The full light curve for 1925 to 1930 in B-band and V-band is displayed in Figure 1. This light curve shows a nova with a flat topped light curve with a peak that is at V=8.5 (and certainly not significantly brighter).

The second refutation is simply that DO Aql brightened to only a visual magnitude of 8.5 in its 1925 eruption, while any prior eruption would only come to the same peak brightness, so a DO Aql event in 5 BC could never have been detected by the Magi or by the Chinese. Kidger reasonably requires that any nova must be very bright, brighter than V=0 or so, to be discovered and considered significant. So we have a huge gap from V=8.5 to V=0.

Kidger was aware of this problem, but he addressed this with only one unsatisfying sentence, "If DO Aquilae really did have a huge eruption in 5 B.C., it is quite possible that this explosion removed the need for a further big eruption of the star for a very long time: successive ones might be smaller until, once more, the system reaches a new and massive crisis." That is, he gives no motivation, theory, or precedent to account for how an ancient eruption could possibly be more than 8.5 magnitudes brighter than in 1925.

Indeed, we have three strong modern astrophysical reasons for knowing that DO Aql could not possibly have had a significantly brighter eruption two millennia ago. First, all eruptions from a single recurrent nova always have the same light curve and reach the same peak magnitude. We know this empirically from my comprehensive reconstruction of all known recurrent nova eruptions[2]. And we know this theoretically because the light curve is determined by only the white dwarf mass, its magnetic field, the composition of the accreted material, and the orbital period (which determines the accretion rate), with none of these changing over the time scale of millions of years. Second, even though I have found the first case where a classical nova event leads to a recurrent nova event[10], classical and recurrent novae both lead to approximately the same peak brightness. This is known both empirically[2,11,12] and theoretically[13]. The average peak absolute magnitude for both is -8.0 mag, with a 1-sigma scatter of 1.3 mag and a total range of -6.1 to -10.7. Third, we know that the 1925 eruption had ordinary spectral, photometric, and color development for an F-class nova, so the peak absolute magnitude would have been approximately -8.0±1.3. With this, the putative 5 B.C. eruption would have to have been at least 8.5 magnitudes brighter to be discovered by the Magi, for a peak absolute magnitude of -16.5±1.3 or more luminous. This explosive energy is that of a Type II supernova, with the implication that any such explosion would have to completely destroyed the system. We still see DO Aql, so the system could not have had any eruption two thousand years ago that destroyed the system. And to emphasize this more, we know that there is no way for a cataclysmic variable with a low mass white dwarf to produce any explosion with supernova energies. From all three reasons, we know with high confidence that DO Aql could not have had an eruption in 5 B.C. that was significantly brighter than its 1925 eruption.

***Third Refutation; Any Nova Is Meaningless To The Magi***

Historically, all the naturalistic explanations for the Star of Bethlehem invoke some astronomical event that would be impressive to modern astronomers. But this completely misses the point, because the only people to attach original meaning to the Star are the Magi, and they were *astrologers*, not astronomers. Astrologers have no way to place a nova onto a horoscope. Astrologers have no interpretation for any nova. (We do know a lot about what the ancient astrologers actually practiced and believed, for example with roughly contemporary books by Ptolemy and Firmicus.) Novae are irrelevant and meaningless to astrologers. As such, any hypothesis of nova-as-Star is nonsensical, because the Magi could not have been motivated to travel to Judea by a nova. This third refutation is as strong as any refutation possible from ancient history.

This third refutation can be extended, with the same power, to *any* nova, supernova, or comet. That is, Persian astrologers had no way to place any nova/supernova/comet onto their horoscopes, nor would they have any way of interpreting any nova/supernova/comet. All novae/supernovae/comets are irrelevant and meaningless to the Magi, and thus they cannot be the Star of Bethlehem.

This third refutation is taken from the 1999 book of Michael Molnar[14], titled "*The Star of Bethlehem: the Legacy of the Magi*". Molnar has argued that the *astrological* orientation of the Magi has ruled out prior naturalistic explanations, because they are all *astronomical* events that excite *modern astronomers*, but not *ancient astrologers*. The bottom line is that no nova, supernova, or comet could ever have motivated the Magi to travel to Judea.

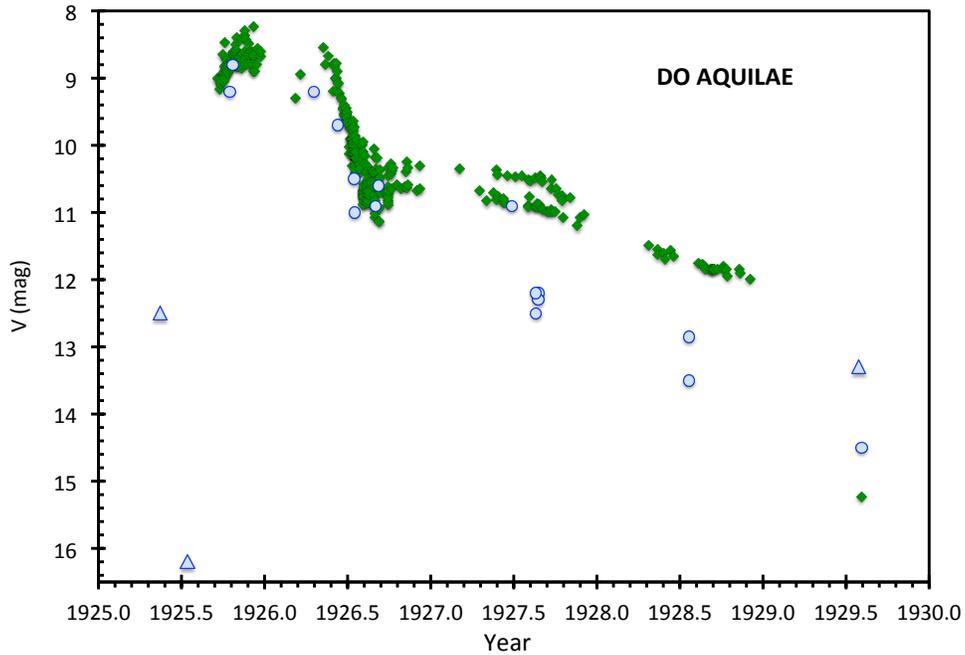

**Fig. I**

This light curve of DO Aql includes all published magnitudes plus my new measures from the Harvard plates. The small (green) diamonds are for V-band measures, the (blue) circles are the B-band magnitudes, and the (blue) triangles are for upper limits in the B-band. The light curve shows a flat topped peak from nearly 1925.8 to 1926.4 at V=8.5. The three magnitudes from 1926.19 to 1926.30 demonstrate that the nova did not have any significant brightening during the gap in observations in the middle of the peak, while the upper limits at 1925.37 and 1925.54 demonstrate that the peak is not just a late plateau for some earlier and much brighter peak. Thus, this light curve proves that DO Aql never got significantly brighter than V=8.5. The two streams of V magnitude around 1927.6 are simply due to observers using slightly different sequences of comparison stars. (This is a ubiquitous problem before the 1970s when dealing with sequences fainter than tenth magnitude). The B-V color starts out around zero, and increases to around one magnitude in the tail, with this being the normal development for novae.